\documentclass[%
 reprint,
 amsmath,amssymb,
 aps,
prb,
 longbibliography
]{revtex4-2}

\usepackage{graphicx}
\usepackage{dcolumn}
\usepackage{bm}
\usepackage{physics}
\usepackage{mleftright}
\usepackage{siunitx}
\usepackage{makecell}
\usepackage{hyperref}

\usepackage{ulem}
\usepackage[usenames]{xcolor}
\usepackage{epstopdf}

\definecolor{mygreen}{RGB}{0,204,102}

\DeclareMathAlphabet{\mathpzc}{OT1}{pzc}{m}{it}

\begin{document}

\preprint{APS/123-QED}

\title{Exchange Interactions in Rare-earth Magnets $A_2$PrO$_3$ ($A$= alkali metals): Revisited}

\author{Seong-Hoon Jang}
\affiliation{
 Institute for Materials Research, Tohoku University, 2-1-1 Katahira, Aoba-ku, Sendai, 980-8577, Japan
}
\author{Yukitoshi Motome}
\affiliation{
 Department of Applied Physics, The University of Tokyo, Tokyo 113-8656, Japan
}

\date{\today}

\begin{abstract}

Rare-earth materials hold promise to realize exotic magnetic states owing to synergy between electron correlations and spin-orbit coupling.
Recently, quasi-two-dimensional honeycomb magnets $A_2$PrO$_3$ ($A$ = alkali metals) were predicted to be good candidates for Kitaev quantum spin liquids, as the Kitaev-type bond-dependent anisotropic interactions dominate over bond-independent isotropic Heisenberg ones.
However, experimental observations are negative, questioning the energy hierarchy in Pr$^{4+}$ ions assumed in the conjecture on the basis of the conventional Russell-Saunders coupling scheme. 
We here revisit the exchange interactions in these Pr compounds, by explicitly taking into account the ionic states beyond the assumption. 
We show that, while increasing the octahedral crystal field splitting, which was assumed to be negligibly small, relative to the spin-orbit coupling, the Kitaev-type interactions are suppressed to be subdominant compared to prevailing Heisenberg ones. 
Our finding compromises the contradiction as arising from the peculiar ionic state of the high valence Pr$^{4+}$.
\end{abstract}

\pacs{Valid PACS appear here}
\maketitle

{\it Introduction.}---The Kitaev model, a pivotal theoretical framework leading to quantum spin liquids (QSLs), arises from frustrated magnetic exchanges on a honeycomb lattice, suppressing conventional magnetic ordering~\cite{KI2006B}. 
The emerging interest lies in the exploration of excited nonlocal quasiparticles inherent within QSLs, as they exhibit resilience against decoherence, thereby offering potential avenues for topological quantum computing~\cite{AN1973, SA1992, KI2006B, NA2008, BA2010, ZH2017A, SA2017}. 
The key feature of this model is bond-dependent Ising-type interactions, dubbed the Kitaev interactions, which arise from the intricate interplay between electron correlation and spin-orbit coupling (SOC). 
This is notably observed in the spin-orbit coupled Mott insulators with the Kramers doublet $\Gamma_7$ described by $j_{\rm eff}=1/2$ pseudospins~\cite{KH2005, JA2009}.
Extensive investigations have focused on the proximity of the $j_{\rm eff}=1/2$ doublet stemming from the low-spin $d^5$ electron configuration under the octahedral crystal field (OCF) in various quasi-two-dimensional honeycomb compounds, including $A_2$IrO$_3$ ($A$=Na, Li), $\alpha$-RuCl$_3$, and related materials~\cite{JA2009, PhysRevB.93.214431, TR2017, WI2017, HE2018, PhysRevMaterials.2.054411, TA2019, MO2020, Motome2020, PhysRevMaterials.5.104409}.
Such investigations were recently extended to the high-spin $d^7$ electron configuration under the OCF, particularly evident in quasi-two-dimensional honeycomb materials featuring Co$^{2+}$ and Ni$^{3+}$ ions~\cite{LI2018, SA2018, Motome2020}.

\begin{figure*}[t!]
\includegraphics[width=1.8\columnwidth]{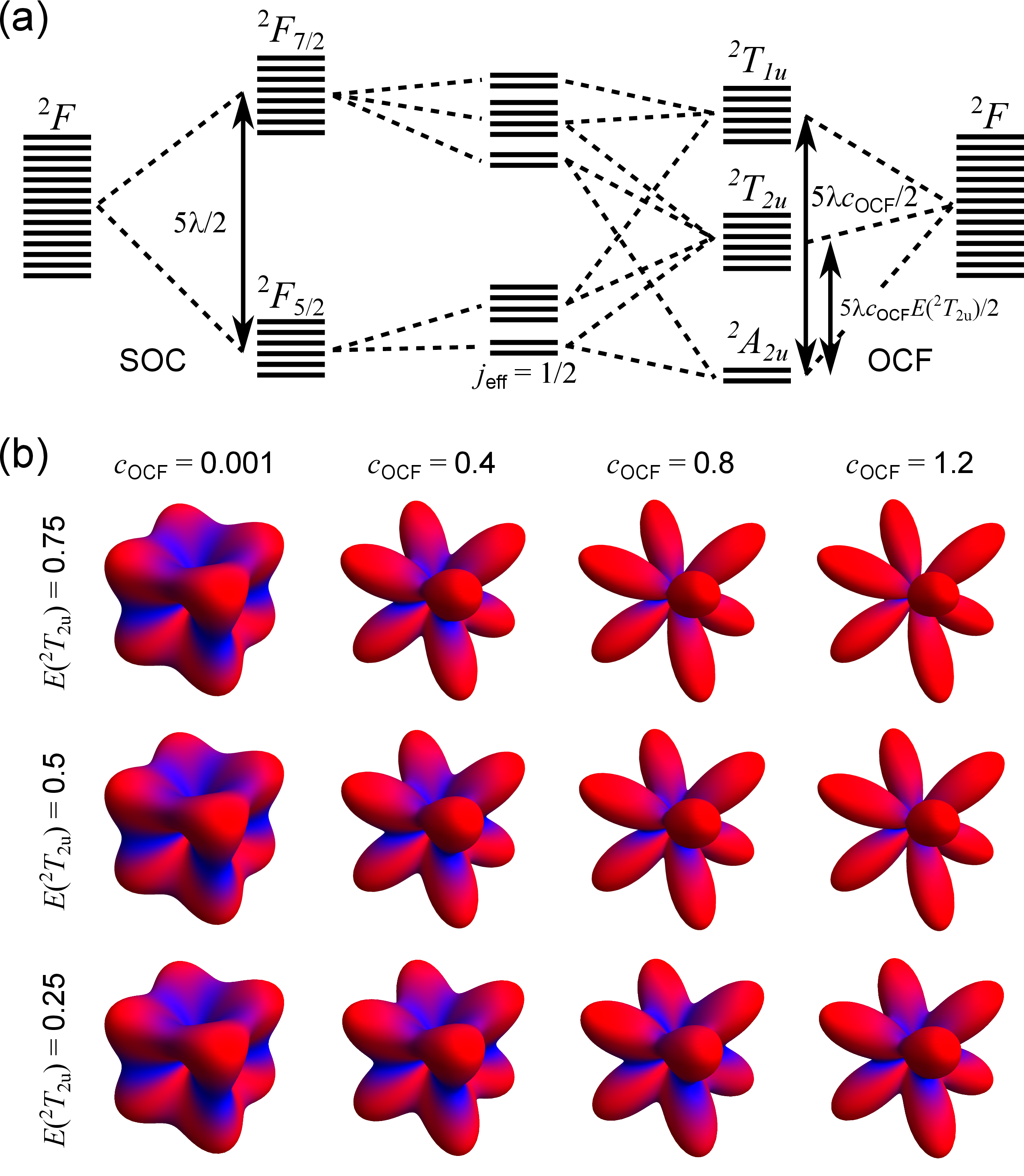}
\caption{\label{fig:Kramers}
(a) Schematic representation of the $14$-fold degenerate $f$-orbital manifold $^2F$ split by the spin-orbit coupling (SOC) effect and the octahedral crystal field (OCF). 
Under the SOC, the energy level difference between the $^2F_{5/2}$ sextet and the $^2F_{7/2}$ octet is given as $5\lambda/2$, where $\lambda$ denotes the SOC coefficient.
Under the OCF, the energy level differences between the $^2A_{2u}$ doublet and the $^2T_{2u}$ sextet and between the $^2A_{2u}$ doublet and the $^2T_{1u}$ sextet are given as $5\lambda c_{\rm OCF}/2$ and $5\lambda c_{\rm OCF} E(^2T_{2u})/2$, respectively; $c_{\rm OCF}$ and $E(^2T_{2u})$ are the two parameters in $\mathpzc{H}_{\textrm{OCF}}$ given as Eq.~(\ref{eq:H_OCF}).
The ground-state Kramers doublet described by $j_{\rm eff}=1/2$ pseudospins is formulated under SOC and OCF.
(b) Wave functions of the ground-state Kramers doublet with red and blue denoting spin-up and spin-down density profiles, respectively, given $c_{\rm OCF}=0.001$, $0.4$, $0.8$, and $1.2$ and $E(^2T_{2u})=0.25$, $0.5$, and $0.75$.
We show the pseudospin-up states only.
} 
\end{figure*}

In advancing materials design for spin-orbit coupled Mott insulators, recent focus has turned to rare-earth quasi-two-dimensional honeycomb materials hosting $f$ electrons.
The possibility of Kitaev interactions was initially examined for double perovskites~\cite{LI2017} and an Yb pyrochlore compound~\cite{RA2018}. 
Subsequently, {\it ab-initio}-based methods have identified quasi-two-dimensional honeycomb magnets $A_2$PrO$_3$ ($A$ = alkali metals) as another candidates with the $\Gamma_7$ doublet in the $4f^1$ electron configuration in Pr$^{4+}$ ions~\cite{JA2019, PhysRevMaterials.4.104420}. 
The estimation suggests that the antiferromagnetic (AFM) bond-dependent anisotropic Kitaev interaction $K$ prevails over the AFM bond-independent isotropic Heisenberg interaction $J$, based on the Russell-Saunders coupling scheme~\cite{RussellSaunders1925} assuming the energy hierarchy $U > \lambda \gg \Delta$; here, $U$, $\lambda$, and $\Delta$ represent the energy scales of Coulomb interaction, SOC, and OCF splitting, respectively.
However, recent experiments indicate that Pr$^{4+}$ ions reside in the intermediate coupling regime $\lambda \sim \Delta$, and $J$ is predominant over $K$~\cite{PhysRevB.103.L121109, Ramanathan2023A, Ramanathan2023B}, challenging this conventional assumption. 
The observed phenomenon can likely be attributed to the unusually high valence $4+$ and the reduced ionic radius $0.85$~\AA, distinguishing it from other instances with conventional trivalent ions, such as Ce$^{3+}$ $1.01$~\AA, Pr$^{3+}$ $0.99$~\AA, Nd$^{3+}$ $0.983$~\AA, and Sm$^{3+}$ $0.958$~\AA~\cite{SH1976}.
This discrepancy challenges theoretical examination in the context of exotic magnetism, especially the Kitaev QSL.
Given the exploration has been extended to other rare-earth magnets such as YbCl$_3$~\cite{PhysRevB.102.014427} and SmCl$_3$~\cite{PhysRevMaterials.6.064405}, reinvestigation of this issue is crucial for future studies of rare-earth Kitaev magnets.

\begin{figure*}[th!]
\includegraphics[width=2.0\columnwidth]{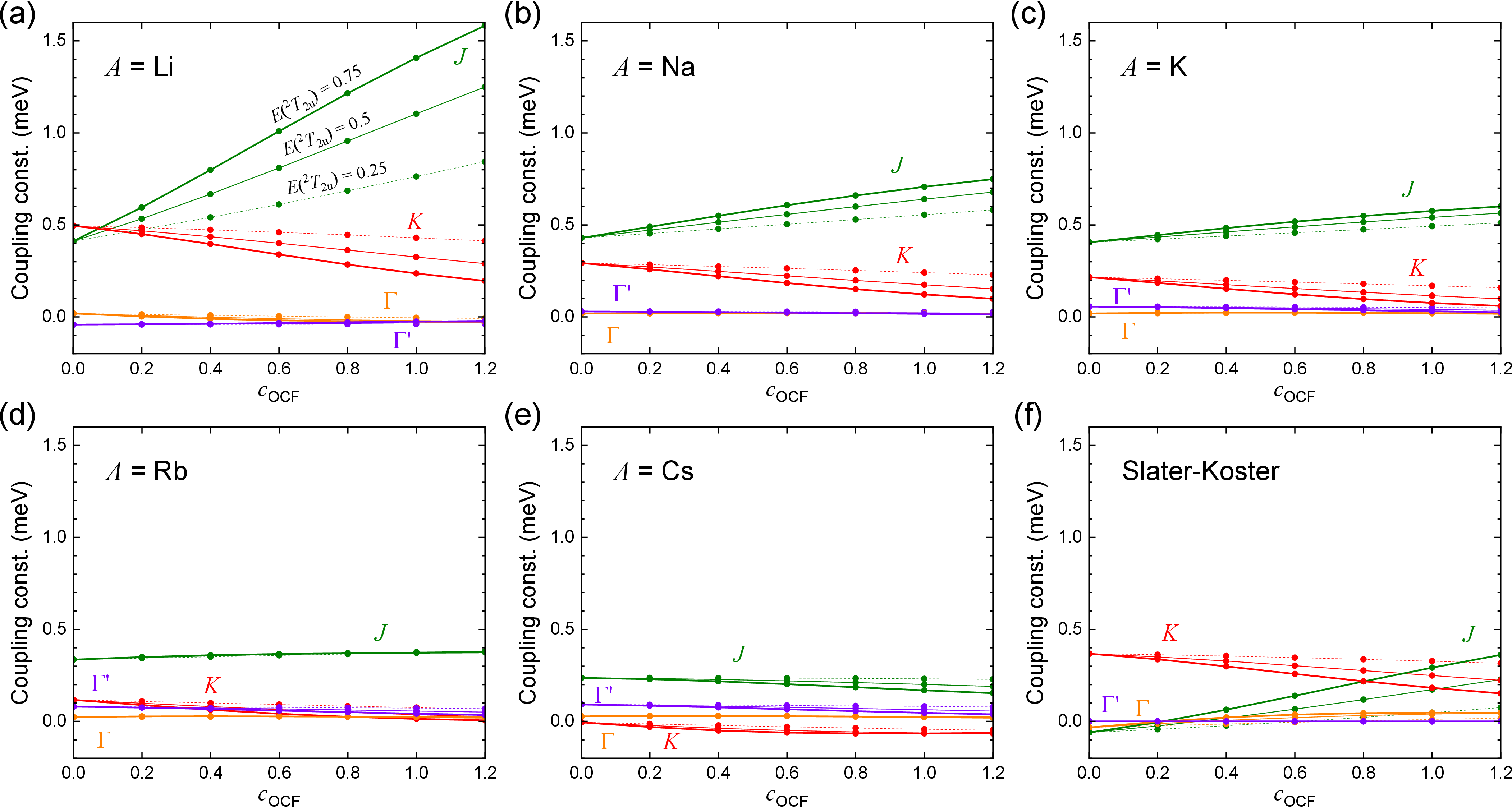}
\caption{\label{fig:couplingconst}
Coupling constants for $A_2$PrO$_3$ with (a) $A$=Li, (b) Na, (c) K, (d) Rb, and (e) Cs as functions of the two parameters $c_{\rm OCF}$ and $E(^2T_{2u})$ in $\mathpzc{H}_{\textrm{OCF}}$ given as Eq.~(\ref{eq:H_OCF}).
The results for the Slater-Koster parameters are also shown in (f) for comparison.
The green, red, orange, and purple lines represent the Heisenberg $J$, Kitaev $K$, and off-diagonal couplings $\Gamma$ and $\Gamma'$, respectively; see Eq.~\eqref{eq:H_extKitaev}.
The bold, solid, and dashed lines represent the results at $E(^2T_{2u}) = 0.75$, $0.5$, and $0.25$, respectively.
The onsite Coulomb interaction $U$, the Hund's-rule coupling $J_{\rm H}/U$, and the SOC coefficient $\lambda$ are set to $6$~eV, $0.1$, and $0.12$~eV, respectively.
} 
\end{figure*}

In this Letter, the low-energy magnetic properties of $A_2$PrO$_3$ is revisited by simultaneously taking into account the electron correlation, SOC, and OCF in the estimation of the exchange interactions. 
We find that in the intermediate regime of $\lambda \sim \Delta$ the AFM $K$ diminishes to a subdominant role, while the AFM $J$ becomes predominant; the other symmetric off-diagonal interactions $\Gamma$ and $\Gamma'$ are relatively small, particularly for $A$=Li, Na, and K.
These results provide an accurate understanding of the magnetic properties of $A_2$PrO$_3$, revealing the complex interaction among electronic correlation, SOC, and OCF due to the high valency of Pr$^{4+}$.

{\it Ground-state Kramers doublet.}---Let us discuss the formation of the $4f^1$ ground-state Kramers doublet in the presence of both SOC and OCF.
The detailed formulation of the SOC contribution in the Hamiltonian, $\mathpzc{H}_{\textrm{SOC}}$, is outlined in the authors' previous work~\cite{PhysRevMaterials.4.104420}, where the SOC coefficient $\lambda$ is specified as $0.12$~eV~\cite{PhysRevMaterials.4.104420}.
As depicted in the left part of Fig.~\ref{fig:Kramers}(a), $\mathpzc{H}_{\textrm{SOC}}$ splits the $14$-fold degenerate $f$-orbital manifold $^2F$ into the $^2F_{5/2}$ sextet and the $^2F_{7/2}$ octet with energy levels denoted as $E(^2F_{5/2})$($=-2\lambda$) and $E(^2F_{7/2})$($=3\lambda/2$), respectively, as shown in the right part of Fig.~\ref{fig:Kramers}(a).
Meanwhile, the contribution from OCF, $\mathpzc{H}_{\textrm{OCF}}$, splits $^2F$ into the $^2A_{2u}$ doublet ($A$ orbital with spin quantum number $\sigma = \pm1$), the $^2T_{2u}$ sextet ($\xi$, $\eta$, and $\zeta$ orbitals with $\sigma = \pm1$), and the $^2T_{1u}$ sextet ($\alpha$, $\beta$, and $\gamma$ orbitals with $\sigma = \pm1$), at the energy denoted as $E(^2A_{2u})$, $E(^2T_{2u})$, and $E(^2T_{1u})$, respectively.
Taking $E(^2A_{2u})=0$, $0 < E(^2T_{2u}) < 1$, and $E(^2T_{1u})=1$, we can define $\mathpzc{H}_{\textrm{OCF}}$ as
\begin{widetext}
\begin{equation}
\begin{split}
\mathpzc{H}_{\textrm{OCF}} = \frac72 \lambda c_{\rm OCF} \Bigg(  & \sum_{f=\alpha, \beta, \gamma} \sum_{\sigma}  c^{\dagger}_{f\sigma}c_{f\sigma} + E(^2T_{2u}) \sum_{f=\xi, \eta, \zeta} \sum_{\sigma}  c^{\dagger}_{f\sigma}c_{f\sigma} - \frac{6 + 6E(^2T_{2u})}{14} \Bigg),
\end{split}
\label{eq:H_OCF}
\end{equation}
\end{widetext}
where $c_{f\sigma}^\dagger$ and $c_{f\sigma}$ denote the creation and annihilation operators of an $f$ orbital with $\sigma = \pm1$, respectively, and the last term offsets the barycenter of the $\mathpzc{H}_{\textrm{OCF}}$-split energy levels to zero.
$c_{\rm OCF}$ is the parameter to control the relative strength of SOC and OCF; at $c_{\rm OCF}=1$, the energy level splittings by $\mathpzc{H}_{\textrm{SOC}}$ and $\mathpzc{H}_{\textrm{OCF}}$ become equal, namely, $E(^2F_{7/2}) - E(^2F_{5/2}) = E(^2T_{1u}) - E(^2A_{2u})$; see Fig.~\ref{fig:Kramers}(a). 
Note that the limit of $c_{\rm OCF} \to 0$ corresponds to the Russell-Saunders coupling scheme assumed in the previous studies~\cite{JA2019, PhysRevMaterials.4.104420}.
Consequently, in the presence of both SOC and OCF, the $\Gamma_7$ Kramers doublet described by the $j_{\rm eff}=1/2$ pseudospin gives the lowest-energy state, as shown in the center of Fig.~\ref{fig:Kramers}(a). 

We illustrate the wave functions of the ground-state Kramers doublet with varying $c_{\rm OCF}$ and $E(^2T_{2u})$ in Fig.~\ref{fig:Kramers}(b): $c_{\rm OCF}=0.001$, $0.4$, $0.8$, and $1.2$ and $E(^2T_{2u})=0.25$, $0.5$, and $0.75$.
In the {\it ab initio} calculations by the authors~\cite{JA2019, PhysRevMaterials.4.104420}, it was estimated that $c_{\rm OCF} \sim 0.2$ and $E(^2T_{2u}) \sim 0.2$.
For small values of $c_{\rm OCF}$, the wave function closely resembles the ideal $\Gamma_7$ state, while as $c_{\rm OCF}$ increases, along with an increment in $E(^2T_{2u})$, it tends to align more closely with the $A$ orbital. 
as notably influenced by the lowest-energy level $E(^2A_{2u})$ as described in Eq.~(\ref{eq:H_OCF}).

{\it Effective exchange couplings.}---The effective Hamiltonian for the $j_{\rm eff}=1/2$ pseudospins described above is given by
\begin{equation}
\begin{split} 
\mathpzc{H} = \sum_\gamma \sum_{\langle i,j \rangle_\gamma} \Bigl\{  & K S_i^\gamma S_j^\gamma + J \mathbf{S}_i \cdot \mathbf{S}_j + \Gamma (S_i^{\gamma'} S_j^{\gamma''} + S_i^{\gamma''} S_j^{\gamma'}) \\
& + \Gamma' (S_i^{\gamma} S_j^{\gamma'} + S_i^{\gamma'} S_j^{\gamma} + S_i^{\gamma} S_j^{\gamma''} + S_i^{\gamma''} S_j^{\gamma}) \Bigr\},
\end{split}
\label{eq:H_extKitaev}
\end{equation}
where $S_i^\gamma$ denotes the $\gamma$ component of the spin-$1/2$ operator $\mathbf{S}_i$ at site $i$ and the sum of $\langle i,j \rangle_\gamma$ is taken for one of three types of nearest-neighbor bonds on the honeycomb structure ($\gamma=x,y,z$); $(\gamma, \gamma', \gamma'')$ are given as $(x, y, z)$ and its cyclic permutations~\cite{JA2009, CH2010, CH2013, RA2014, RU2019}.

For the estimates of exchange coupling constants, we perform perturbation calculations in terms of the hopping term $\mathpzc{H}_{\textrm{hop}}$. 
For the intermediate states with $4f^2$ electron configuration in the perturbation, we consider the Coulomb interaction term $\mathpzc{H}_{\textrm{int}}$ in addition to $\mathpzc{H}_{\textrm{SOC}}$ and $\mathpzc{H}_{\textrm{OCF}}$. 
The detailed formulations are outlined in the authors' previous work~\cite{PhysRevMaterials.4.104420}.
For $\mathpzc{H}_{\textrm{int}}$, we set the onsite Coulomb interaction $U$ and the Hund's-rule coupling $J_{\rm H}$ to $U=6$~eV and $J_{\rm H}/U=0.1$, respectively.
Then, the $4f^2$ intermediate states are also parametrized by $c_{\rm OCF}$ and $E(^2T_{2u})$.
For $\mathpzc{H}_{\textrm{hop}}$, we incorporate the effects of indirect $4f$-$2p$-$4f$ and direct $4f$-$4f$ electron hopping processes both, as determined by using {\it ab initio} calculations for $A_2$PrO$_3$ with $A$=Li, Na, K, Rb, and Cs provided in the authors' previous works~\cite{JA2019, PhysRevMaterials.4.104420}, alongside the Slater-Koster transfer integrals $t_{pf\sigma} = 0.7$~eV, $t_{pf\pi}/t_{pf\sigma} = -0.7$, $t_{ff\sigma} = 0.1$~eV, $t_{ff\pi}/t_{ff\sigma} = -0.5$, $t_{ff\delta}/t_{ff\pi} = -0.5$, and $t_{ff\phi}/t_{ff\delta} = -0.5$ for comparison~\cite{TA1980}.

Figure~\ref{fig:couplingconst} summarizes the results for the coupling constants $J$, $K$, $\Gamma$, and $\Gamma'$ while varying $c_{\rm OCF}$ for $A_2$PrO$_3$ ($A$=Li, Na, K, Rb, and Cs) and the Slater-Koster scenario. 
In the case of $A$=Li, the magnitude of AFM $J$ exhibits a monotonic increase with $c_{\text{OCF}}$, while the AFM $K$ monotonically decreases [see Fig.~\ref{fig:couplingconst}(a)]. 
We note that, at $c_{\text{OCF}} = 0.001$, the estimates are given as $J=0.412$~meV, $K=0.496$~meV, $\Gamma=0.0190$~meV, and $\Gamma'=-0.0414$~meV; these are similar to the previous ones for the ideal $\Gamma_7$, that is, $J=0.339$~meV, $K=0.585$~meV, $\Gamma=0.0133$~meV, and $\Gamma'=-0.0459$~meV, where $\mathpzc{H}_{\textrm{int}}$ and $\mathpzc{H}_{\textrm{SOC}}$ are considered not simultaneously but sequentially and $\mathpzc{H}_{\textrm{OCF}}$ was neglected in the intermediate states~\cite{JA2019}.
The AFM $J$ becomes predominant over the AFM $K$, particularly at large $c_{\text{OCF}}$ alongside large $E(T_{2u})$, while $\Gamma$ and $\Gamma'$ are comparatively small across variations in $c_{\text{OCF}}$ and $E(T_{2u})$.
The result indicates that in the intermediate regime $\lambda \sim \Delta$ the system is approximately described by the AFM Heisenberg model with subdominant AFM Kitaev interactions.

Similar trends are observed for $A$=Na and K, where the disparities between AFM $J$ and AFM $K$ are diminished [see Figs.~\ref{fig:couplingconst}(b) and \ref{fig:couplingconst}(c)].
In the case of $A$=Rb, the effective interactions between pseudospins are predominantly described by AFM $J$, with negligibly small contributions from $K$, $\Gamma$, and $\Gamma'$; $J$ remains nearly constant ($\sim0.4$~meV) across variations in $c_{\text{OCF}}$ and $E(T_{2u})$ [see Fig.~\ref{fig:couplingconst}(d)].
For $A$=Cs, a transition occurs where $K$ becomes ferromagnetic (FM), while AFM $J$ and positive $\Gamma'$ compete with FM $K$ [see Fig.~\ref{fig:couplingconst}(e)].
In the Slater-Koster scenario, a trend akin to that observed for $A$=Li, Na, and K is noted, whereas the AFM $K$ remains larger than the AFM $J$ in a wider range of $c_{\text{OCF}} < 0.8$ [see Fig.~\ref{fig:couplingconst}(f)].
Overall, a significant change in the dominance between AFM $J$ and AFM $K$ becomes evident as $c_{\text{OCF}}$ increases, possibly attributable to the increased contribution from the direct hopping between the $A$ orbitals.
This contrasts with the trend observed in the previous works, where AFM $K$ would prevail over AFM $J$, assuming negligibly small $\mathpzc{H}_{\textrm{OCF}}$~\cite{JA2019, PhysRevMaterials.4.104420}.

We estimate the realistic values of coupling constants in the regime of $\lambda \sim \Delta$, assuming $c_{\rm OCF}=1.0$ and $E(T_{2u})=0.25$, $0.5$, and $0.75$; see Table~\ref{tab:Values} in Appendix~\ref{sec:svcc}. 
Notably, for $A$=Na and $E(T_{2u})=0.75$, it is demonstrated that $J=0.707$~meV, $K=0.122$~meV, $\Gamma=0.0183$~meV, and $\Gamma'=0.0169$~meV.
The provided values are indeed comparable to one set of fitting values derived from inelastic neutron scattering data of Na$_2$PrO$_3$: $J=1.1$~meV and $K=0.20$~meV~\cite{PhysRevB.103.L121109}.
It is noteworthy that these values are considered large for rare-earth-based materials, but well accounted for by our realistic estimates. 
This highlights the significance of our calculation scheme in understanding the magnetic behavior of $A_2$PrO$_3$.

{\it Summary.}---We have investigated the low-energy magnetic properties of $A_2$PrO$_3$ by explicitly considering the interplay between SOC and OCF. 
We found that in the regime where SOC and OCF are comparable, the anisotropic AFM Kitaev interaction $K$ diminishes to a subdominant role, while the isotropic AFM Heisenberg interaction $J$ becomes predominant. 
The presented values for $J$ and $K$ exhibit a notable sensitivity to the parameters $c_{\text{OCF}}$ and $E(T_{2u})$, indicating the intricate nature of the magnetic interactions in $A_2$PrO$_3$.
Additionally, we showed that the symmetric off-diagonal interactions $\Gamma$ and $\Gamma'$ are comparatively small for several alkali metal ions ($A$=Li, Na, and K). 
We also provided realistic values, which are comparable to fitting parameters obtained from recent experimental data~\cite{PhysRevB.103.L121109}, further corroborating the validity of our theoretical framework.
Overall, our findings offer reliable estimates of the magnetic properties of $A_2$PrO$_3$ with high valence Pr$^{4+}$, emphasizing the importance of considering both SOC and OCF effects in the analysis of rare-earth magents.

{\it Acknowledgments.}---The authors thank R. Okuma for informative discussions. 
This work was supported by by JST CREST (Grant No.~JP-MJCR18T2), and JSPS KAKENHI Grant Nos.~19H05825 and 20H00122.


\appendix
\renewcommand\thetable{\thesection\arabic{table}}

\section{Specific values of effective exchange coupling constants} 
\label{sec:svcc}

In Table~\ref{tab:Values}, we present the coupling constants $J$, $K$, $\Gamma$, and $\Gamma'$ for $A$=Li, Na, K, Rb, and Cs and the Slater-Koster scenario at $c_{\rm OCF}=1.0$ and $E(T_{2u})=0.25$, $0.5$, and $0.75$.
In most cases, AFM $J$ predominates over AFM $K$, particularly evident for $E(T_{2u})=0.75$.

\begin{widetext}
\begin{table*}[h!]
\centering
\caption{\label{tab:Values}Estimates of the isotropic Heisenberg interaction $J$, anisotropic Kitaev interaction $K$, and symmetric off-diagonal interactions $\Gamma$ and $\Gamma'$ for $A$=Li, Na, K, Rb, and Cs and the Slater-Koster scenario at $c_{\rm OCF}=1.0$ and $E(T_{2u})=0.25$, $0.5$, and $0.75$.
The onsite Coulomb interaction $U$, the Hund's-rule coupling $J_{\rm H}/U$, and the SOC coefficient $\lambda$ are set to $6$~eV, $0.1$, and $0.12$~eV, respectively.
}
\begin{ruledtabular}
\begin{tabular}{cccccc}
$A$&$E(T_{2u})$&$J$ (meV)&$K$ (meV)&$\Gamma$ (meV)&$\Gamma'$ (meV)\\
\hline
Li&$0.25$&$0.763$&$0.430$&$-0.00457$&$-0.0388$\\
&$0.5$&$1.10$&$0.326$&$-0.0205$&$-0.0321$\\
&$0.75$&$1.41$&$0.237$&$-0.0253$&$-0.0251$\\
\hline
Na&$0.25$&$0.555$&$0.241$&$0.0217$&$0.0275$\\
&$0.5$&$0.640$&$0.175$&$0.0213$&$0.0221$\\
&$0.75$&$0.707$&$0.122$&$0.0183$&$0.0169$\\
\hline
K&$0.25$&$0.493$&$0.169$&$0.0231$&$0.0510$\\
&$0.5$&$0.541$&$0.115$&$0.0228$&$0.0405$\\
&$0.75$&$0.576$&$0.0760$&$0.0196$&$0.0305$\\
\hline
Rb&$0.25$&$0.373$&$0.0749$&$0.0275$&$0.0718$\\
&$0.5$&$0.376$&$0.0364$&$0.0268$&$0.0563$\\
&$0.75$&$0.373$&$0.0137$&$0.0228$&$0.0418$\\
\hline
Cs&$0.25$&$0.231$&$-0.0424$&$0.0294$&$0.0815$\\
&$0.5$&$0.201$&$-0.0629$&$0.0276$&$0.0632$\\
&$0.75$&$0.169$&$-0.0655$&$0.0230$&$0.0464$\\
\hline
Slater-Koster&$0.25$&$0.048$&$0.327$&$0.0110$&$0$\\
&$0.5$&$0.173$&$0.249$&$0.0400$&$0$\\
&$0.75$&$0.292$&$0.183$&$0.0476$&$0$\\
\end{tabular}
\end{ruledtabular}
\end{table*}
\end{widetext}

\bibliography{main}

\end{document}